# Size dependent Acoustic Phonon Dynamics of $CdTe_{0.68}Se_{0.32}$ Nanoparticles in Borosilicate glass

Sanjeev K. Gupta and Prafulla K. Jha[a]
Nanotechnology Centre, Department of Physics, Bhavnagar University, Bhavnagar, 364 002, India.

A. K. Arora
Materials Science Division, Indira Gandhi Centre for Atomic Research,
Kalpakkam, 603102, India.

## Abstract

Low frequency acoustic vibration and phonon linewidth for $CdTe_{0.68}Se_{0.32}$ nanoparticle embedded in borosilicate glass are calculated using two different approaches by considering the elastic continuum model and fixed boundary condition. The presence of medium significantly affects the phonon peaks and results into the broadening of the modes. The line width is found to depend inversely on the size, similar to that reported experimentally. The damping time and quality factor have also been calculated. The damping time that is of the order of picoseconds decreases with the decrease in size. High value of quality factor for $l$=2 normal mode suggests the less loss of energy for this mode.

a) prafullaj@yahoo.com

## I. Introduction

Semiconductor nanoparticles or quantum dots have been extensively studied in the past several years due to their wide range of application and an important model system in the study of three dimensional quantum confinements.[1] As size is reduced of a nanoparticle, the electronic structure is modified as compared to the bulk with the oscillator strength concentrated into a few discrete transitions.[2,3] Recently, the semiconductor nanoparticles have received attention due to its importance in biological applications as a replacement for the dye molecules.[4,5] In addition, the semiconductor nanoparticles doped glasses have long attracted attention for their potential uses in optoelectronic devices. Since the glass matrix is transparent in visible range, the optical properties of these composite materials are determined by those of the nanoparticles. As the emission of phonon is one of the most important electronic dephasing mechanisms, there has been much interest in the vibrational properties of semiconductor nanoparticles. Apart from modifying the electronic structure, confinement also affects coupling of electrons to the optical and acoustic phonons. In the case of nanoparticle, the coupling is enhanced due to the localized acoustic vibrational motion. In addition, the low frequency phonon modes (acoustic phonons) of nanoparticles bear a unique signature of their structural and mechanical properties directly reflecting the impact of confinement on the ionic movement. This is in contrast to the optical frequency domain whose features are determined by the electronic response. Optical phonons,[6-15] confined acoustic phonons,[8,9-13] surface[16,17] and disorder-activated phonons[18,19] have been extensively investigated recently by both theoretical and experimental means. The acoustic phonons are found to be accountable for the width of the lowest exciton transition in nanocrystals.[20,21] These transitions in nanoparticles often have a linear temperature dependence attributed to the thermal excitation of acoustic phonons.[22]

The acoustic phonon sidebands encountered in the photoluminescence (PL) are found to affect energy relaxation in nanoparticles. In a recent paper, Bragas et al[23] performed the spontaneous Raman scattering (RS) and pump-probe measurements for $CdTe_{0.68}Se_{0.32}$ quantum dots embedded in borosilicate glass matrix and ascribed them to the spheroidal $l=0$, 2 and torsional $l=1$ modes of a spherical particle though the torsional $l=1$ mode is not a Raman active mode and should be absent in Raman scattering.[24] The $CdTe_{0.68}Se_{0.32}$ nanoparticle is of special importance due to its high quantum efficiencies. Moreover, by changing the size of the $CdTe_{0.68}Se_{0.32}$ nanoparticle, it can be made to radiate throughout the visible spectrum. Bragas et al[23] found that the frequency of the symmetric acoustic mode ($l=0$), only one to be experimentally observed in the standard pump-probe experiments exhibits a strong dependence on the laser energy in the pump-probe but not in the RS. However, there is a large variation in the frequency of fully symmetrical mode (SPH, $l=0$) from these two experiments. To further correlate the observed oscillations with the nanoparticle vibrations, they also calculated theoretically the low frequency modes for $l=0$ and $n=0$, $l=2$, $n=0$ by using the classical Lamb's model with free boundary condition,[25] which ignores the presence of borosilicate glass. In addition, they used the same ratio of sound velocities $V_l/V_t = 2.3$ as of CdSe for the calculation of the low frequency spheroidal modes. This value of $V_l/V_t$ is quite right for the CdSe but for CdTe it is about 25% more. Therefore the ratio of sound velocities $V_l/V_t$ used for $CdTe_{0.68}Se_{0.32}$ by Bragas et al[23] in their calculation is higher and lead to the higher value of frequency for the phonon modes.[14] While they obtained relatively good agreement with this higher and incorrect value of $V_l/V_t$ for the $l=0$ and 2 modes with Raman data for a QD radius equal to the average radius of the distribution, the difference in calculated and experimental data was quite large for other sizes particularly for $l=0$ mode. However, the theoretical data agree reasonably well with the pump-probe data for $l=0$ mode particularly near $1/R \approx 0.2$ nm$^{-1}$ i.e. for larger nanoparticle. But the agreement becomes poor when correct velocities are used. They have also obtained the full-width at half maximum (FWHM) of both Raman and pump-probe spectra with size of the $CdTe_{0.68}Se_{0.32}$ nanoparticles, which show large difference. However, they have not compared these data with the theoretical calculation, not possible due to the constraint of the model. Therefore, the size dependence of the mode line width is not known theoretically. Since, the $CdTe_{0.68}Se_{0.32}$ nanoparticle is embedded in the borosilicate glass the effect of glass matrix on the confined acoustic phonon modes can not be ignored and hence the free sphere model used by them is not applicable in the present case. It is important to consider the effect of the matrix materials on these frequencies as they become complex valued due to the damping of the vibrational modes. The energy of the modes is mechanically radiated away from the nanoparticle into the surrounding glass matrix[26] results into the damping of modes. Therefore, it is of interest to calculate the low frequency modes of $CdTe_{0.68}Se_{0.32}$ embedded in borosilicate glass with the fixed boundary conditions applied to Lamb's solution and compare the results with the experimentally obtained data as well as with the Lamb's solution known as free sphere model (FSM). This will not only make actual comparison but also will be helpful in understanding the actual role of borosilicate glass on the low frequency modes and coherent acoustic phonon modes[28] observed in the experimental study.

In the present paper, we report a theoretical calculation of the low frequency acoustic phonon modes (LFM) for the $CdTe_{0.68}Se_{0.32}$ nanoparticles embedded in borosilicate glass by applying the approaches that also considers the interaction of the nanoparticles with the surrounding borosilicate glass. These approaches are the complex frequency model (CFM)[16,28] and core shell model (CSM)[29-31] and have been used to describe the low frequency acoustic phonon modes for several systems. In addition, the parameters such as damping time, linewidth quality factor etc. associated with the damping of acoustic oscillation resulting due to the presence of the surrounding medium are also calculated. The paper is organized as follows: in sec II, we describe the methodology to obtain the CFM and CSM models; in section III, we present results and discussion followed by the conclusion in section IV.

## II. Methodology and Computation

In the present study, Lamb's theory of vibrations of a homogenous, free-standing elastic body of spherical shape is used for the calculation of the low frequency phonon modes of borosilicate embedded $CdTe_{0.68}Se_{0.32}$ nanoparticles. However, to understand the key features, the vibrations of the $CdTe_{0.68}Se_{0.32}$ nanoparticle embedded in borosilicate glass are calculated by using three different methods based on classical Lamb's theory, in the present paper: (i) Eigen value equation of free isotropic sphere (Lamb's model)[25] (ii) Complex frequency model applicable to embedded nanoparticle[28-29] and (iii) Core-shell model applicable to both free as well embedded nanoparticle.[31] The displacement field $\bar{u}(\bar{r},t)$ of an elastic medium of density $r(\bar{r})$ is governed by Navier's equation

$$c_{ijkl,j} u_{k,l} + c_{ijkl} u_{k,lj} = r \ddot{u}_i \qquad (1)$$

Where $c_{ijkl}(\vec{r})$ is the fourth rank elastic constant tensor field. However, the present equation of motion for a spherical elastic body is written as[25,26]

$$(\lambda + 2\mu)\vec{\nabla}(\vec{\nabla}\cdot\vec{u}) - \mu\vec{\nabla}\times(\vec{\nabla}\times\vec{u}) = \rho\ddot{u} \quad (2)$$

Where, $\lambda$ and $\mu$ are Lame's constants and $\rho$ is the mass density of nanoparticle. Introducing proper scalar and vector potential, the solution of Eq. (2) gives two vibrational modes, namely a spheroidal mode and a torsional mode. The eigenvalue equations for spheroidal modes were derived as[11,26]

$$2\left[\eta^2 + (l-1)(l+2)\left\{\frac{\eta j_{l+1}(\eta)}{j_l(\eta)} - (l+1)\right\}\right]\frac{\zeta j_{l+1}(\zeta)}{j_l(\zeta)} - \frac{1}{2}\eta^4 + (l-1)(2l+1)\eta^2 + [\eta^2 - 2l(l-1)(l+2)]\frac{\eta j_{l+1}(\eta)}{j_l(\eta)} = 0 \quad (3)$$

Where $\eta$ is the dimensionless eigenvalues and $j_l(\eta)$ or $j_l(\zeta)$ are the spherical Bessel's function of first kind and $\zeta = \eta(V_t/V_l)$. The frequencies of these modes are indexed with integers $n$ and $l$, known as the harmonic numbers and angular quantum number respectively. The Raman peak frequencies of the spheroidal mode can be obtained from the expression as[26]

$$\eta_l^S = \frac{\omega_l^S R}{V_t} \quad (4)$$

Where $V_l$ and $V_t$ are the average longitudinal and transverse velocities. The torsional modes are purely transverse in nature and independent of material property and are defined for $l \geq 1$ and are orthogonal to the spheroidal modes.[24] The spheroidal modes are characterized by $l \geq 0$, where $l=0$ is the symmetric breathing mode, $l=1$ is the dipolar mode and $l=2$ is the quadrupole mode. Duval[24] has shown that for a spherical particle only the $l=0$ and 2 modes are Raman active. The $l=0$ mode is of particular interest as this is observable in both pump-probe and Raman experiments.[33]

In the complex frequency model (CFM),[28-31] the nanoparticle is surrounded by a homogeneous and isotropic matrix of density $\rho_m$ and speeds of sound $V_{lm}$ and $V_{tm}$. The CFM is the result of fixed boundary condition and is different from the free boundary condition in the sense that it takes into account the stresses at the nanoparticles boundary, continuity of ū and force balance at the nanoparticle-matrix interface due to the existence of the boundaries between particle and matrix interface. The boundaries between the particle and matrix modify the confined phonon modes and even some times responsible for the appearance of new modes. The boundary condition at large $R$ is that ū is an outgoing travelling wave. This model yields complex frequency, the real part $Re(\omega)$ represents the frequency of the free oscillations. The imaginary part $Im(\omega)$, which is always positive, stands for the attenuation coefficient of the oscillations because the amplitude of the displacement is proportional to $e^{iRe(\omega)t-Im(\omega)t}$ at any place and therefore $Im(\omega)$ yields damping. The real frequency, $Re(\omega)$ in the present case are shifted from the frequency $\omega$ obtained from Lamb equations.[28] Therefore, the resulting complex nature of the frequency due to radiative decay enables to estimate both frequencies and their damping. Unlike the free vibrations of an elastic body the scalar and vector potentials include first kind of spherical Hankel functions, whose variable is purely imaginary in the region outside the nanoparticle in embedded systems. Verma et al[9] have also used a similar approach, to calculate the low frequency phonon modes of nanoparticle embedded in matrix. This approach also yields complex frequency similar to the present CFM and show the shifting of frequency in the presence of surrounding medium. In experimental observations of such modes a broadened peak is observed with its centre at real frequency.

Though CFM model is able to predict the real frequency i.e the actual peak position and broadening (FWHM) of the peak due to the damping actually observed in the Raman spectra but contains several inherent limitations. In these the most important one is the absence of valid wave functions required for the Raman scattering spectra calculations. In addition, the correspondence to quantum theory is also unclear due to the modes not being orthonormalizable and blowing up exponentially with the radial co-ordinate.[35] To overcome this, Portalés et al[32] proposed an approach namely core-shell model (CSM) considering nanoparticle as core surrounded by a macroscopically large spherical matrix as shell which leads to real valued mode frequencies.[31] Therefore for calculating the absolute intensity or shape of a low frequency Raman spectrum or for the comparison with the experimental Raman spectra, the CSM is the right approach.

In CSM, the motion of nanoparticle, the mean square displacement $\langle u^2 \rangle_p$ which is nothing but the measure of the internal motion of the nanoparticle is obtained as a function of mode's frequency and can be expressed as[31]

$$\langle u^2 \rangle_p = \frac{1}{v_p} \int_{R<R_p} \|\bar{u}(\bar{R})\|^2 d^3\bar{R} \qquad (5)$$

Where, $v_p$ is the volume of the particle. The mean-square displacement $\langle u^2 \rangle_p$ is the key ingredient for the calculation of phonon spectrum. The plots of $\langle u^2 \rangle_p$ with mode frequency are continuous and peaks in the plots correspond closely to the real part of the CFM. Half widths at half maximum of these peaks correspond closely to the imaginary parts of CFM frequencies.[28] The mean square displacement can be compared to the Raman spectra, as the Raman intensity depends on the square of the displacement. The general idea behind this core-shell model approach (CSM) is that the motion of the nanoparticle is smaller than the wavelength of the acoustic phonon modes of matrix materials. Moreover, the Raman spectrum is governed by both $\langle u^2 \rangle \tilde{n}_p$ and the electron-phonon interaction matrix element. The matrix element which appears as multiplicative constant, determines the overall intensity while $\langle u^2(w) \rangle \tilde{n}_p$ determines both amplitude and the spectral line shape. Therefore, a preliminary step for obtaining the Raman spectrum would be to study the amplitude of nanoparticles vibrations as a function of phonon frequency. In a recent study, Groose and Zimmermann[36] have also calculated the energy dependence of the phonon amplitude for the similar system with the inclusion of the effect of strain arising due to the mismatch at the interface of nanoparticles and matrix. They found that the calculated phonon amplitude is proportional to a quantity which modifies the electron-phonon interaction and therefore will be responsible for the change in the profile of the Raman spectra.[33]

## III. Results and Discussion

In the following, we present results on the investigation of low frequency vibrational modes of $CdTe_{0.68}Se_{0.32}$ nanoparticle embedded in borosilicate glass ($CdTe_{0.68}Se_{0.32}$-Borosilicate) by using three different models described above. A low frequency analysis for $CdTe_{0.68}Se_{0.32}$-Borosilicate is presented here for a range of particle sizes. The eigenfrequencies for the selected spheroidal modes have been obtained by using both CFM and CSM and are presented in table I. For the sake of comparison the table also includes the frequencies obtained using Lamb's theory (Eq. (4)) and the reported experimental Raman and pump-probe data[23] on these modes. Apparent good agreement with Lamb's model was found by Bragas et al[23] as incorrect sound velocities were used. Agreement becomes poor when correct sound velocities are used. The sound velocities for these calculations are estimated by the linear interpolation between the velocities of sound for CdTe and CdSe obtained by using the elastic constants.[38] The sound velocities so obtained for the CdTe$_{0.68}$Se$_{0.32}$ are $3.219x10^5$ cm/s and $1.764x10^5$ cm/s for $V_l$ and $V_t$ respectively. Fig. 1 presents the size dependence of the spheroidal ($l$=0, 1, 2 and $n$=0) modes for the free boundary condition approach (Lamb's theory). The inverse size dependence of the frequency of low frequency phonon modes is evident. It is seen from the Fig. 1 and table I that the frequency for a nanoparticles of particular size is minimum for the $l$=2 and $n$=0 mode. Of interest are the major differences in the acoustic phonon frequency for stress free and borosilicate-environment cases particularly with the correct sound velocities, the spheroidal modes for free and embedded nanoparticle are presented in Fig. 2. The borosilicate-environment produces significant modifications in the acoustic phonon energies. This result is important for the understanding of the appearance of side bands in PL spectrum and optical emission spectrum of CdTe$_{0.68}$Se$_{0.32}$ nanocrystals.[23] A reduction in the frequency of the phonon modes is clearly visible which is due to the energy loss of phonon mode during the process of energy radiation by the modes away from the CdTe$_{0.68}$Se$_{0.32}$ nanoparticle into the borosilicate glass. This clearly reflects the importance of considering the surrounding matrix to describe the phonon spectra of CdTe$_{0.68}$Se$_{0.32}$ nanoparticle embedded in borosilicate glass.

The variation of mean square displacement $\langle u^2 \rangle_p$ with frequency for the low frequency phonon modes (SPH $l$=0, 1 and TOR $l$=1, 2) obtained from CSM are presented for the 3.9 nm CdTe$_{0.68}$Se$_{0.32}$-Borosilicate nanoparticle in Fig. 3. $\langle u^2 \rangle_p$ is plotted individually for each mode verses the mode's energy in separate panel. To make it easier to compare $\langle u^2 \rangle_p$, only one peak is presented for SPH, $l$=0. The oscillating behavior of the phonon amplitude resembles the eigenmodes of an isolated sphere.[36] The $\langle u^2 \rangle_{\tilde{n}_p}$ is the main ingredient for the calculation of Raman spectra and therefore the peak positions in the plot can be compared with the peak positions or Raman shift in experimental Raman spectra and with real frequency obtained from CFM model. Groose and Zimmermann[36] have also shown that the calculated phonon amplitude depend on a quantity which modifies the electron-phonon interaction and hence the Raman spectra through electron-phonon matrix element. This fact suggests that the preliminary step in the calculation of Raman spectra is the study of the amplitude of nanoparticle vibrations as a function of phonon energy. The frequency dependence of $\langle u^2 \rangle_{\tilde{n}_p}$ for $l \geq 1$ spheroidal modes shows the mixture of longitudinal and transverse waves traveling at different speeds. For the spheroidal modes, there are actually two amplitudes inside the sphere, otherwise there is a single amplitude inside the sphere.[25,27,29,31] These amplitudes are complex numbers, however only the absolute values are plotted. The shift of the peak to higher frequency for smaller particle size (not shown here) is similar to that found in the Lamb and CFM models. Raman scattering detects the fluctuations in polarizability tensor $a$. The polarizability tensor $a$ oscillates at the same frequency as that of the mechanical vibrations of the $CdTe_{0.68}Se_{0.32}$ nanoparticle. As far as the effect of surrounding medium is concerned the resulting spectra show the homogeneously broadened peak of the phonon modes. This broadening of the peak in the spectra is caused by the particle interaction with the surrounding matrix and mismatch in the acoustic impedance of the $CdTe_{0.68}Se_{0.32}$ nanoparticle and borosilicate glass, as the low frequency phonon modes vibration involves the whole system, the nanoparticle, and surrounding medium. Though it is possible to incorporate QD size distribution in the calculations, it is not done in the present study as the same is not known for the samples used by Bragas et al.[23] Furthermore, they probed size selected particles in the same samples using resonance tuning of electronic excitation.

Figure 4 presents the low frequency acoustic phonon modes estimated theoretically by using CSM for $l=0$ and $l=2$ spheroidal modes as a function of inverse of the size ($1/R$) for $CdTe_{0.68}Se_{0.32}$ nanoparticle embedded in borosilicate glass. For comparison the present figure also includes the data from experimental pump-probe and Raman measurements.[23] One can see that while the Raman data follows both $l=0$ and $l=2$ modes the pump-probe data follows only $l=0$ mode. This is due to the fact that standard pump-probe configuration detect only the fully symmetric modes such as $l=0$ mode.[33] One can see that the frequencies obtained for $l=2$ matches reasonably well with the experimental Raman data ($l=2$) while for $l=0$, it is close to the pump-probe data in particular near $1/R = 0.2$ nm$^{-1}$. The poor agreement of calculated and experimental Raman data, particularly for $l=0$ seems to be due to the large error bars in experimental Raman data. In view of the large scatter in the data ($l=0$) the inverse size dependence is not clear. Moreover, the usual trend of frequency variation with size is only seen for resonantly selected radius in pump-probe measurement. The present CSM calculation is performed for the embedded nanoparticles while the theoretical calculation performed in ref.[23] is for free boundary condition (Lamb's equation). Therefore, present calculations are expected to make better representation of the vibrational mode characteristic of $CdTe_{0.68}Se_{0.32}$ nanoparticle embedded in borosilicate glass. The reduction in the value of frequencies in the present calculations (for both $l=0$ and $l=2$) with respect to the earlier calculated values[23] and present Lamb's solution is due to the inclusion of surrounding medium which gives rise to complex frequency.

The present paper also reports the size variation of phonon linewidth for low frequency acoustic phonon modes of $CdTe_{0.68}Se_{0.32}$ nanoparticle embedded in borosilicate glass. The calculation of line width gives not only the real test of the model but also gives the estimates of damping of vibrational modes. The Fig. 5 shows the variation of phonon mode linewidth with the size for the $CdTe_{0.68}Se_{0.32}$ nanoparticle obtained from both CFM and CSM approaches for $l=0$ spheroidal modes. For comparison the present figure also includes the experimental phonon linewidth observed from the pump-probe and Raman scattering measurements.[23] The linewidth calculated from the CSM are comparable with the linewidth obtained from Raman scattering.[23] However, for Raman data the $1/R$ dependence is not clear due to the large error bars, but the pump-probe data show clear $1/R$ dependence for the available data. It is seen from figure 5 that the calculated linewidth obtained in the present study and that found in the Raman measurement is larger than the pump-probe data. This is due the fact that the acoustic phonon signal originating in pump-probe is from smaller nanoparticles and very narrow distribution of nanoparticles size.[23] The present calculations use the average size of the nanoparticles in the distribution. The Fig. 5 reveals that the values of linewidth show increase with inverse size ($1/R$).

In figure 6 the variation of damping time which is inverse of damping constant for $l=0$ and 2 spheroidal modes with the size for the $CdTe_{0.68}Se_{0.32}$ nanoparticle embedded in the borosilicate glass is presented. The damping time for normal modes is obtained from the expression $t_D = -1/Im(w)$. Where $Im(\omega)$ the imaginary frequency of phonon modes is obtained from the complex eigenvalue. The torsional $l=1$ mode could not ideally be expected to be Raman active. This could be important in the case of non-spherical nanoparticles. Calculations of strain dependent acoustic phonons for InAs sphere embedded in GaAs material show the similar trend for energy dependence of phonon amplitudes.[36] The oscillating behavior of phonon amplitude here resembles the eigenmodes of an isolated sphere[36] and peaks may be the overtones, which are not investigated in the present study. The Fig. 6 reveals that the damping time for $CdTe_{0.68}Se_{0.32}$ nanoparticle is of the order of *ps* and decreases for the smallar nanoparticles. This implies that the energy is transferred faster to the surrounding medium in the case of smaller particle, which is consistent with other results.[31] In addition to the role of surrounding medium there may be a contribution from the surface of the nanoparticles to the damping of the phonon modes as the damping constant (opposite of damping time) variation with size of the nanoparticle is similar to that of the size dependency of the surface atoms (surface/volume ratio) for a nanoparticle. On the basis of these two effects it appears that the damping of the phonon modes occurs through surface mechanical coupling of the vibrations of nanoparticle to the matrix and therefore the consideration of surrounding medium in the description of vibrations of borosilicate embedded $CdTe_{0.68}Se_{0.32}$ nanoparticle is important. The damping time can be used to determine the quality factor. The quality factor is a quantity which measures sharpness of the response of a resonating system to external excitation and can be expressed as $Q = t_D w$; where $t_D$ is damping time and $w$ is the real frequency obtained in CFM approach. The values of quality factor $Q$ for the 3.9 nm $CdTe_{0.68}Se_{0.32}$ are 1.982 and 0.634 for $l=0$ and $l=2$ phonon mode respectively which is independent of size for any particular mode. This suggests that there is less loss of energy for the $l=0$ mode. It is found that the quality factor $Q$ is always lower for matrix-embedded nanoparticles and independent of the size,[12] This independent behavior of $Q$ on the size may have some other applications.[30-31] The energy loss, which determine $Q$ for a particular mode, also determine the maximum amplitude of the oscillation when the resonance condition is exactly satisfied, as well as the width of the resonance i.e. how far off the resonant frequency the system can be driven and still yield significant oscillation amplitude.

## IV. Conclusion

Low frequency acoustic phonon modes have been calculated for the $CdTe_{0.68}Se_{0.32}$ nanoparticle embedded in the borosilicate glass by using the two models namely the complex frequency approach and core shell model which considers the presence of surrounding medium based on the classical Lamb's theory. These models establish a relation between the particle size, the frequencies, and widths of confined acoustic phonons. A reasonable well agreement between theoretical and experimental results is obtained. The phonon energies are size dependent and are lower for embedded nanoparticles. It is observed that the phonon line width varies similar to the phonon frequency with size of the particle for low frequency phonon modes. The reduction in size as well as the surrounding medium both contributes to the phonon line broadening. The damping time and quality factor are also calculated. The damping time is of the order of picoseconds, which decreases with the decrease in size. The high value of quality factor for $l$=2 mode suggests less loss of energy for this normal model.

**Acknowledgement**

We appreciate comments and suggestions made by referee which greatly improved this paper. The financial assistance from the Department of Atomic energy (DAE-BRNS) is highly appreciated.

**Figures Caption**

1. Spheroidal (S) mode frequency for free $CdTe_{0.68}Se_{0.32}$ nanoparticles versus 1/R using Lamb's model ( _ .. _ .. _ ) for $l=0$, $n=0$; ( _ _ _ ) for $l=1$, $n=0$ and (──) for $l=2$, $n=0$.

2. Effect of borosilicate glass on the lowest spheroidal mode ($l=2$, $n=0$) (──) for Lamb's solution, (------) for CSM and (— — —) for CFM.

3. Mean square displacement of spheroidal and torsional phonon modes for embedded $CdTe_{0.68}Se_{0.32}$ nanoparticles obtained using CSM, (──) and (-------) represent longitudinal and transverse waves respectively.

4. Acoustic mode frequencies for borosilicate embedded $CdTe_{0.68}Se_{0.32}$ nanoparticles versus 1/R, CSM represents calculated values as the result of the core-shell model (Raman Spectra). Raman and pump-probe data from Ref. 23 are also shown.

5. Dependence of the full width at half maximum (FWHM) on 1/R of the $l=0$ and $n=0$ spheroidal phonon mode. Raman and pump-probe data from Ref. 23 are also shown.

6. Size dependency of the Damping time for $l=0$, $n=0$ (-----) and $l=2$, $n=0$ (──) spheroidal modes of embedded $CdTe_{0.68}Se_{0.32}$ nanoparticle.

**Table**

I. Eigen frequencies (in cm$^{-1}$) of CdTe$_{0.68}$Se$_{0.32}$ -Borosilicate nanoparticle of size 3.9 nm.

| Spheroidal Modes | | | | | |
| --- | --- | --- | --- | --- | --- |
| (Eigen frequencies (cm$^{-1}$)) | | | | | |
| $l$ | $n$ | Lamb | CFM | CSM | Experiment |
| 0 | 0 | 27.59 | 26.65 | 26.15 | 32.0$^a$, 24.0$^b$ |
|   | 1 | 48.55 | 52.07 |   |   |
|   | 2 | 74.20 | 80.01 |   |   |
| 1 | 0 | 15.32 | 10.39 | 9.68 |   |
|   | 1 | 30.70 | 26.64 |   |   |
|   | 2 | 34.75 | 37.51 |   |   |
| 2 | 0 | 11.57 | 11.13 | 15.23 | 15.0$^a$ |
|   | 1 | 21.73 | 26.07 |   |   |
|   | 2 | 37.09 | 39.26 |   |   |

$^a$Experimental values (Raman) from Ref. [23].

$^b$Experimental values (Pump-Probe) from Ref. [23].

**Figure 1**

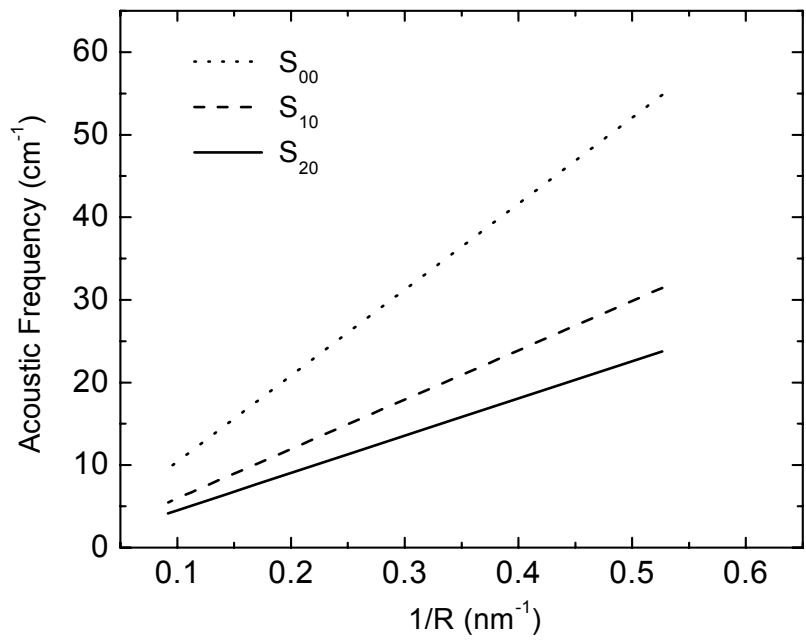

**Figure 2**

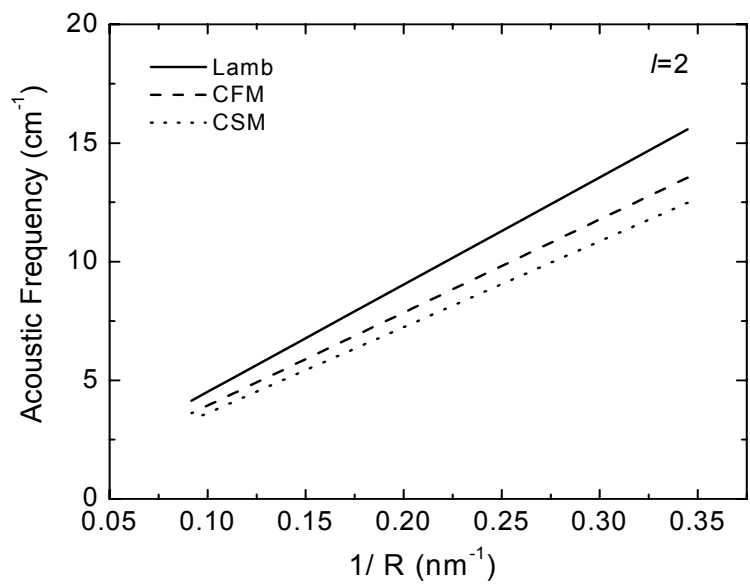

**Figure 3**

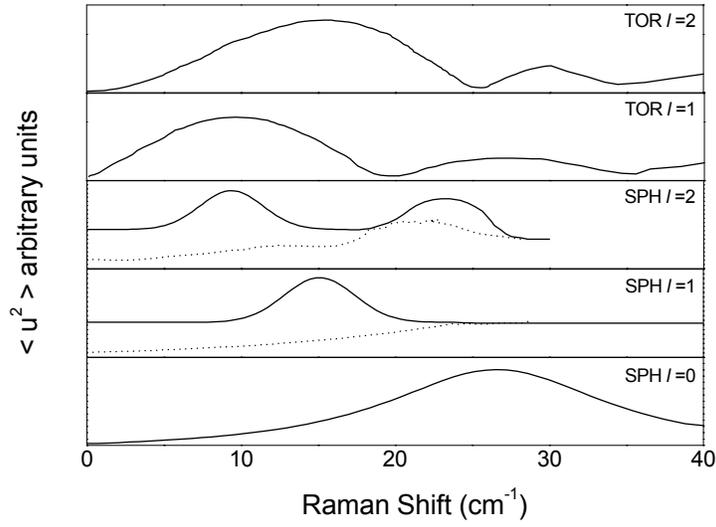

**Figure 4**

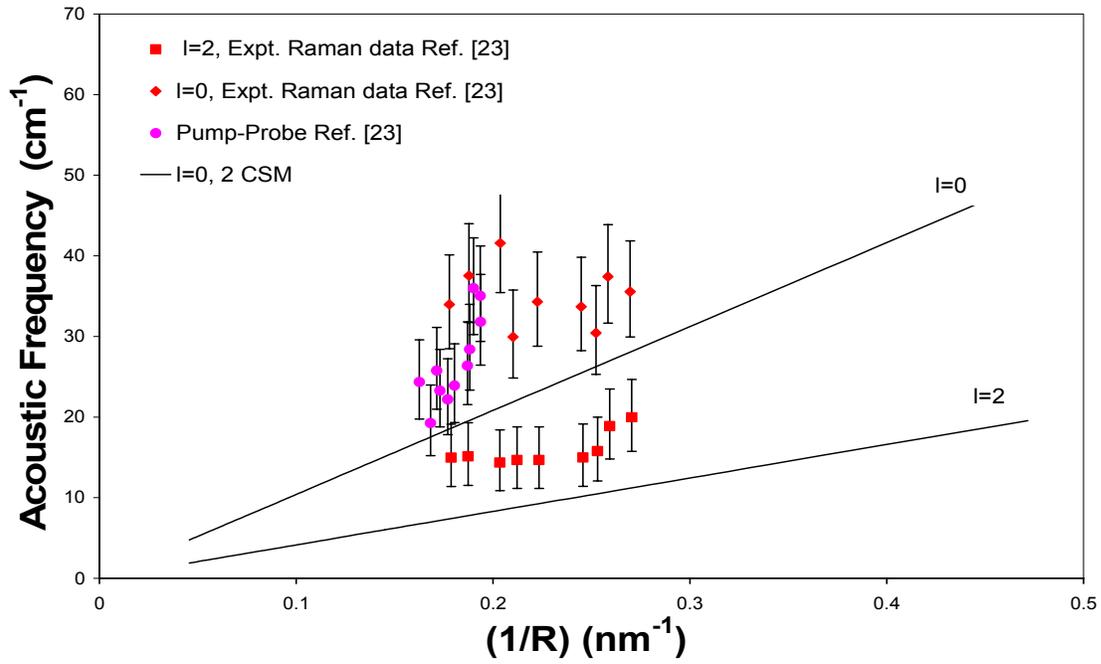

**Figure 5**

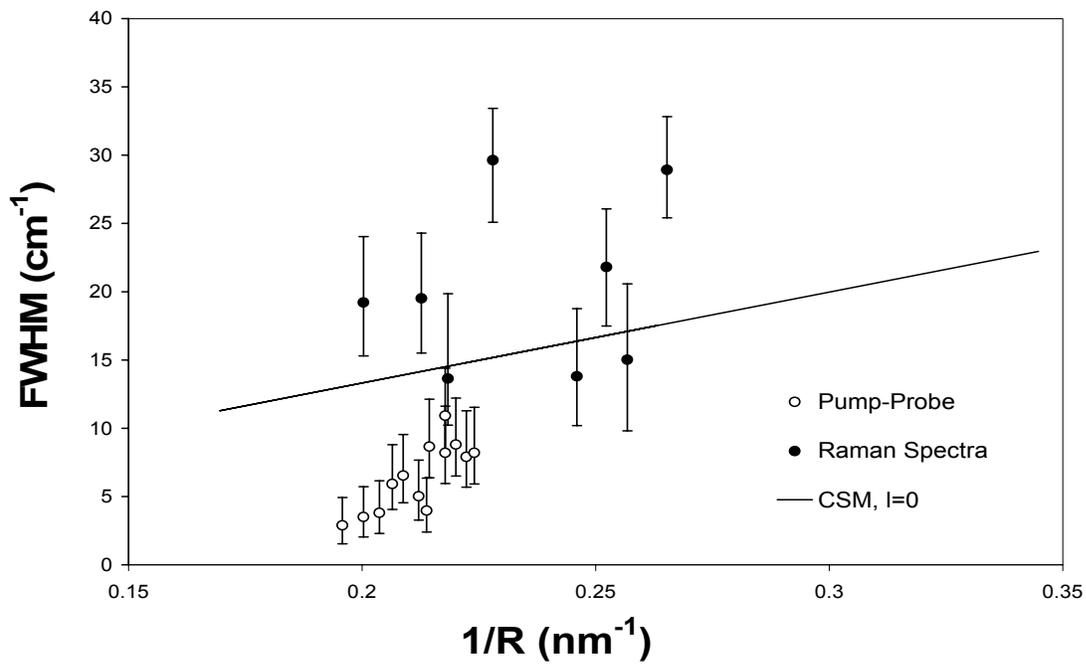

**Figure 6**

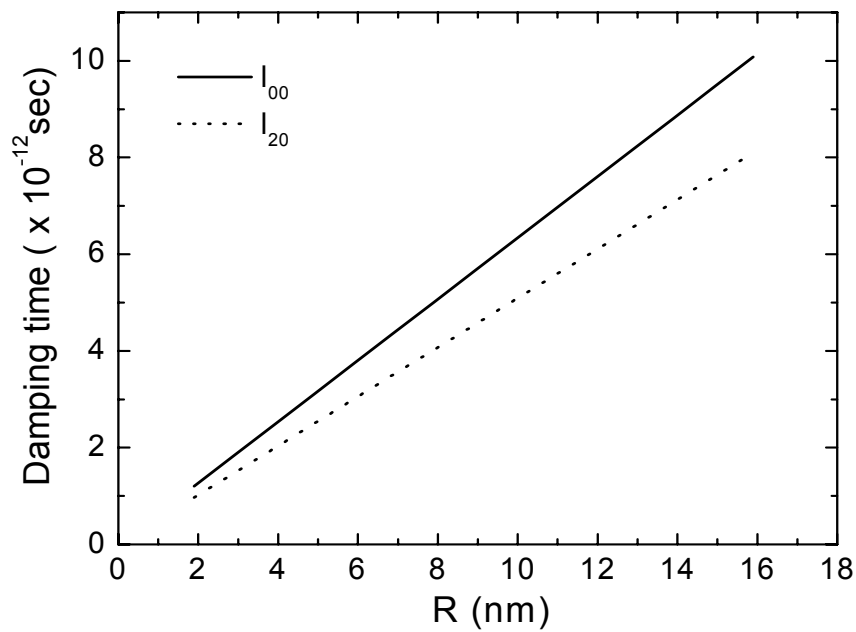